Hidden costs of La Mancha's production model and drivers of change

Máximo Florín Beltrán and Rafael Ubaldo Gosálvez Rey
Regional Centre for Water Studies, University of Castilla-La Mancha

Introduction

The territory of La Mancha, its rural areas, and its landscapes suffer a kind of atherosclerosis ("the silent killer") because of the increase in artificial surfaces, the fragmentation of the countryside by various infrastructures, the abandonment of small and medium-sized farms and the loss of agricultural, material, and intangible heritage.

At the same time, agricultural industrialization hides, behind a supposed productive efficiency, the deterioration of the quantitative and qualitative ecological status of surface and groundwater bodies, and causes air pollution, greenhouse gas emissions, loss of soil fertility, drainage and plowing of wetlands, forgetfulness of the ancestral environmental heritage, of the emergence of uses and customs of collective self-government and reduction of the adaptive capacity of traditional agroecosystems.

This work aims, firstly, to shed light on the true costs of the main causes of environmental degradation in the territory of La Mancha, while deteriorating relations between rural and urban areas and determining the loss of territorial identity of La Mancha. the population. In addition, drivers of change toward a more sustainable social, economic, hydrological, environmental, and cultural production model are identified.

Hidden costs

Unsustainable land use, excessive tillage, land use change, sealing of fertile soils, and other actions release a much higher amount of $CO_2$ into the atmosphere than in other productive sectors. Hence, the transition from land considered rural to artificial soil is one of the most serious environmental problems we face.

The hydrological motto of the current agricultural model of La Mancha is, literally, "Water for a tube!", already in an undisguised way, because using the Tablas de Daimiel National Park as a hostage and infiltration pond, its rescue is claimed in the form of transfers from the Tajo-Segura aqueduct through the so-called Pipeline to the La Mancha Plain, which was conceived exclusively for the supply of drinking water. This point has been reached because 1) The water saved with new irrigation techniques does not compensate for the increase in irrigated area (Figure 1), and 2) Water returns to water bodies by infiltration are not considered.

Food sovereignty has become a heretical litany, in relation to the original meaning of the concept, albeit with several melodies:

- Industrial food production with higher consumption of water resources and waste generation is often much higher than consumption, for example, up to 4.5 times more in the case of pork.
- More and more crops inappropriate for regional soil and climate conditions are expanding, which requires stressing the system with more water, fertilizers, and phytosanitary products than necessary in indicated areas or favoring large losses due to ordinary weather situations, such as frost or drought (Figure 2).
- Surplus harvests with deficient water resources and/or serious environmental impacts are frequent.

- The case of the transformation vineyard: vines are irrigated on trellises, even with more "saving" irrigation techniques, to export alcohol, or worse, burn it.

The European Union's Common Agricultural Policy (CAP), which was supposed to act as a driver of change, does not fix the population, the population is increasingly aging, because young people have difficulty receiving CAP aid, and a huge volume of funds go to large investors, and not to small and medium-sized farmers.

This situation is aggravated by the fact that agri-food cartels impose very low prices at source and small and medium-sized farms are becoming less profitable.

The resulting land centricity causes cascading impacts, for example:
- Margins, banks, and "lost" spaces disappear, affecting the flora and fauna that inhabit them.
- The higher yield you get once the process is finished is often due to the increased use of chemicals.
- Drainage works: i) Water tables fall globally, due to the proliferation of barely controlled extractions for irrigation, and ii) The network of meadows and wetlands is altered by desiccation (Bernáldez et al. 1989).
- All this facilitates the conversion of the entire territory into a cultivated area: i) The streams are rectified, empotrando between crops, ii) The ponds are blinded, or they are excavated to measure using earth dams, iii) The stabling and intensive production of fodder by irrigation and silage are generalized.

The system of tillage of the land does not often cause the maintenance or formation of fertile soils. As specialists in agronomic engineering, they presented at the joint symposium of the Spanish Society of Forest Sciences and the Spanish Association of Terrestrial Ecology that was held in Alcázar de San Juan in 2018, among the bad practices that most reduce soil fertility are 1) over-plowing to prevent any ruderal plant from appearing and 2) not respecting the boundaries between crop fields. In addition, serious processes of alteration of water flows and loss of soil take place.

The use of pesticides acts by reducing the biodiversity of organisms beneficial to soil fertility and promoting populations of pesticide-resistant agricultural pests, which do not encounter competition due to the deterioration of the ecological balance of soils.

As for fertilizers, the effects of chemicals should be added to those of slurry, chickens, and manure, when the interest of their recovery transcends traditional agricultural practices because it is a question of getting rid of residues from industrial meat production.

In the long term, the use of fertilizers can increase the organic matter content of the soil (but also decrease), total nitrogen, and nutrient availability for plants, but it also clearly decreases the C/N ratio and pH due to the continued excessive application of fertilizers.

And all this without considering that agribusiness, which makes intensive use of pesticides and fertilizers, displaces agricultural practices that have contributed to the formation and maintenance of fertile soils, which threaten them, and that the environmental impacts of pesticides and fertilizers, and the health and environmental traits of pesticides incur in the externalization of economic and environmental costs (Figure 3), which calls into question the supposed greater fertility of the soils managed with these practices.

In this sense, the European Commission (2022) has denounced in court that not all the necessary mandatory measures have been foreseen in the Castilla-La Mancha action program and neither have additional measures or reinforced actions been adopted in relation to eutrophication, asregards nitrate pollution, despite showing an upward trend in the pollution of the measuring points of Vulnerable Areas by nitrates, it failed to fulfil its obligations under Articles 3(4) and 5(4) (read in conjunction with Annexes II and III) and 5 of Council Directive 91/676/EEC (1), of 12 December 1991 on the protection of waters against pollution caused by nitrates from agricultural sources.

Drivers of change

The proposed Intervegas Law and the next European Soil Protection Directive (European Parliament 2021) should be highlighted.

The first contemplates (Red Intervegas, 2022), for example:

1) That fertility and the healthy state of soils, without being the only one, constitute a basic prerequisite for the fulfillment of a series of environmental and social objectives and functions. Good soil status is essential for obtaining food, biomass (energy), fiber, fodder, and other products, as well as for ensuring the provision of essential ecosystem services in all regions of the world.
2) Living soils have been built over thousands of years of natural action and through very complex and slow processes. The conjunction of the action of air, water, and living beings gives rise to organic soil, which is part of the existential foundations of humanity. In barren territories, without living soil and without fertility, humanity has no future.
3) Soil is also an essential element of the climate system, constituting the second largest source of carbon storage after the oceans. The protection of soils, their sustainable use, and restoration actions aimed at restoring their fertility are crucial both to mitigate climate change and to adapt to its effects.
4) For all these reasons, it concludes that it is essential to protect and maintain fertile soils so that agriculture can be developed with the guarantee of the permanence of its soil base in all territories, from peri-urban to deeply rural, thus ensuring a production anchored in the territory, close or close, which allows populations, Both urban and rural, can be supplied with fresh, quality fruit and vegetable products associated with Mediterranean and local diets, reducing transport costs and greenhouse gas emissions associated with food that has traveled long distances before reaching our table. The fertile soil is, therefore, an essential strategic food reserve for the support of local agri-food policies, and to favor both the local economy and the maintenance of agricultural landscapes and where educational activities and outdoor enjoyment can be developed.

In any case, it is essential to address in a systemic and synergistic way the change of agricultural model, as a response to the systemic and synergistic nature of the processes and mechanisms of degradation of fertile soils triggered by the current one:

1) Within the actions aimed at mapping fertile soils for their protection, incorporate GIS layers of water resources, in order to be able to make water balances that should be an agricultural reconversion in favor of water sustainability. The big question to answer would be: how many hectares of irrigated land and what characteristics must be eliminated to achieve water sustainability?
2) As social, economic, and perhaps even ecological corrective and compensatory measures, promote production and consumption in proximity and sustainability, progressively eliminating all types of subsidies, but betting on a fair price for producers and the development of brands associated with the integrity of water landscapes and aquatic ecosystems, within a new agricultural model that relies on the green, rural, cultural, heritage, gastronomic, oenological tourism, etc., seeking to fix population with interconnected economic diversification and wealth distribution strategies, and abandoning the current trend towards speculation, agribusiness, industrial meat production, massive energy plants, intensification of logging, and mining driven by volatile markets of raw materials for the development of Information and Telecommunications Technologies, renewable energies (rare earth elements, tungsten, etc.) and industrial fertilizers.

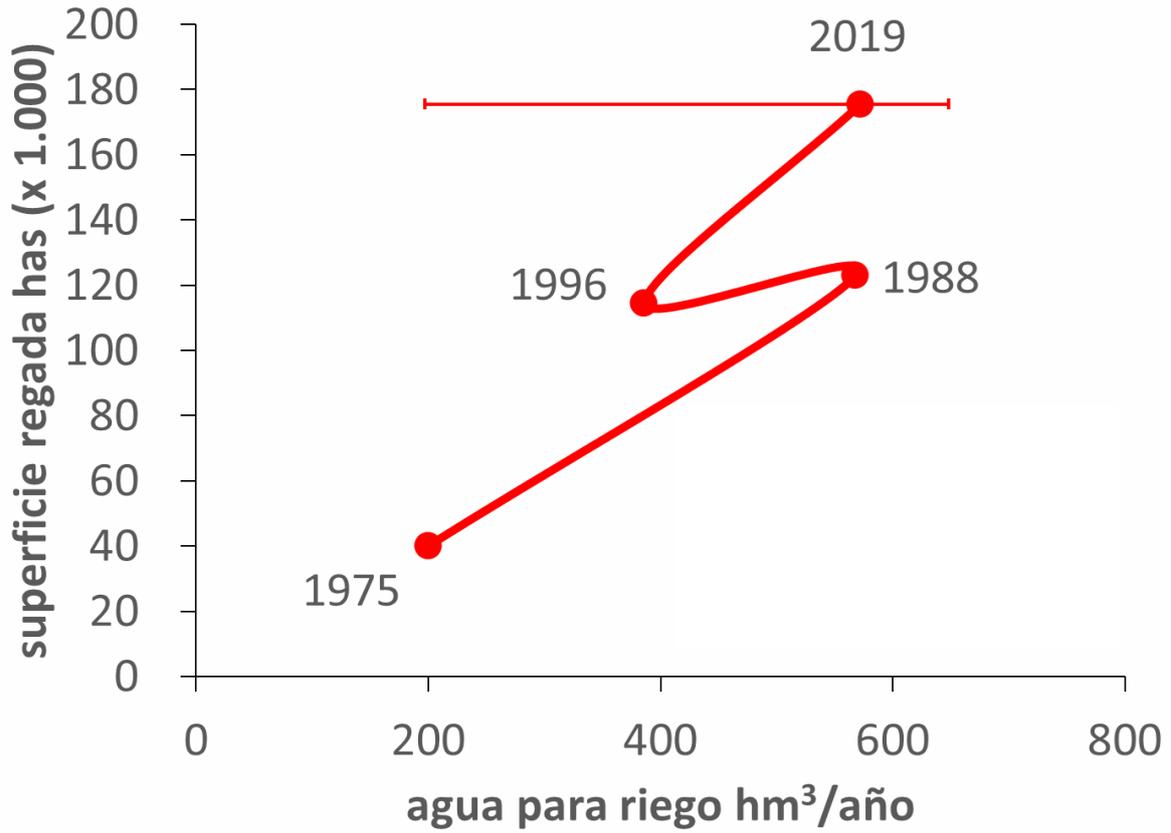

Figure 1. Evolution of irrigation in the Western Mancha aquifer. Source: Guadiana Hydrographic Confederation and Groundwater User Communities.

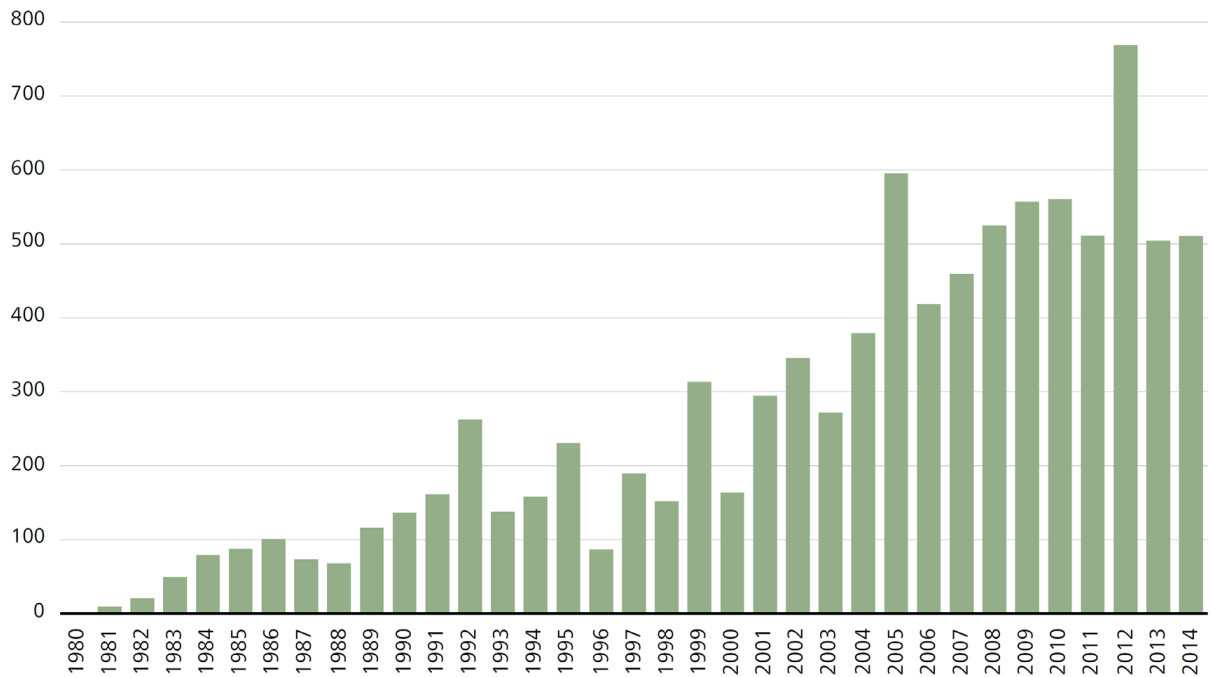

Figure 2. Evolution of agricultural accidents (value in millions of euros). Source: Consorcio de Compensación de Seguros.

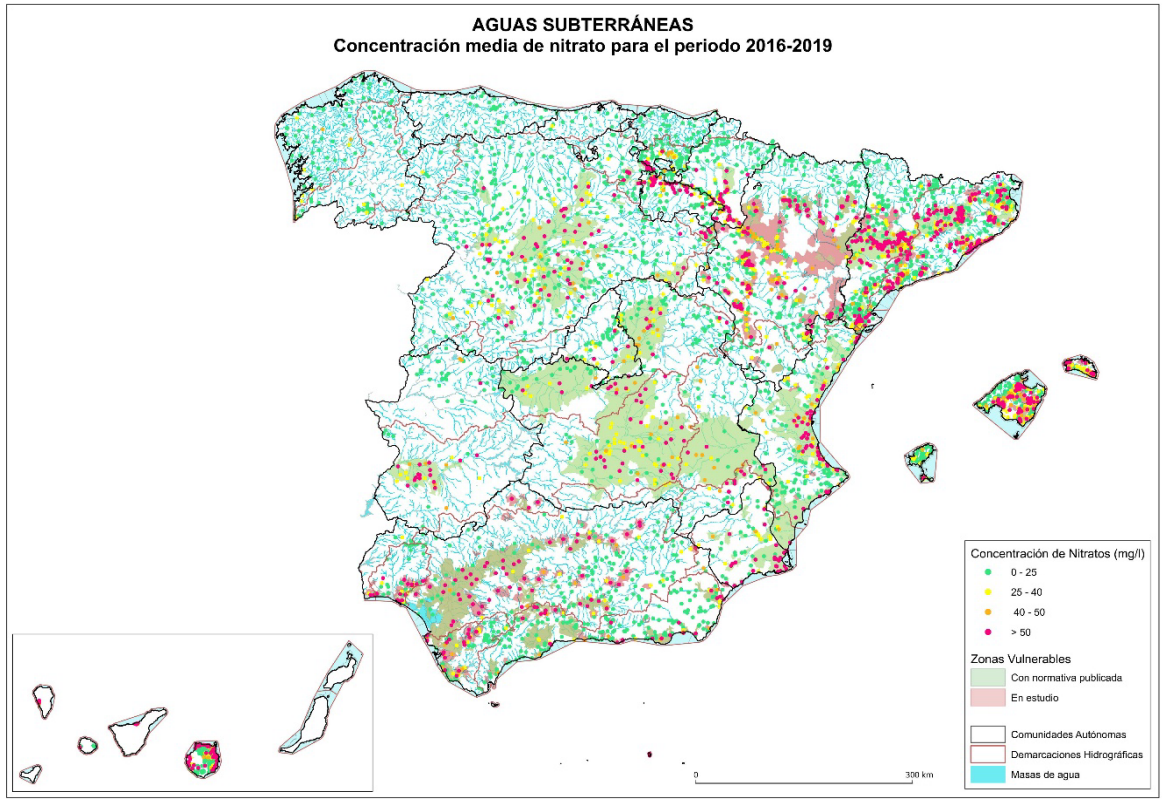

Figure 3. Nitrate Vulnerable Zones. Source: Ministry of Ecological Transition and Demographic Challenge.